\documentclass[11pt, a4paper]{article}
\usepackage{jheppub}
\usepackage[utf8]{inputenc} 
\usepackage[T1]{fontenc}    
\usepackage{url}            
\usepackage{booktabs}       
\usepackage{amsfonts}       
\usepackage{nicefrac}       
\usepackage{microtype}      
\usepackage{lipsum}
\usepackage{multirow}
\usepackage{subfigure}
\usepackage{listings}

\title{
Unsupervised Outlier Detection in Heavy-Ion Collisions
}

\author[a]{Punnathat Thaprasop,}
\author[b]{Kai Zhou,}
\author[b]{Jan Steinheimer}
\author[a]{and Christoph Herold}

\affiliation[a]{School of Physics and Center of Excellence in High Energy Physics \& Astrophysics, Suranaree University of Technology, Nakhon Ratchasima 30000, Thailand}
\affiliation[b]{Frankfurt Institute for Advanced Studies, Giersch Science Center, D-60438 Frankfurt am Main, Germany}
\emailAdd{herold@g.sut.ac.th}

\abstract{
We present different methods of unsupervised learning which can be used for outlier detection in high energy nuclear collisions. The UrQMD model is used to generate the bulk background of events as well as different variants of outlier events which may result from misidentified centrality or detector malfunctions. The methods presented here can be generalized to different and novel physics effects. To detect the outliers, dimensional reduction algorithms are implemented, specifically the Principle Component Analysis (PCA) and Autoencoders (AEN). We find that mainly the reconstruction error is a good measure to distinguish outliers from background. The performance of the algorithms is compared using a ROC curve. It is shown that the number of reduced (encoded) dimensions to describe a single event contributes significantly to the performance of the outlier detection task. We find that the model which is best suited to separate outlier events requires a good performance in reconstructing events and at the same time a small number of parameters.} 


\begin{document}

\maketitle

\flushbottom

\section{Introduction} \label{introduction}
Heavy ion collisions at relativistic beam energies are an abundant source of particles created by the strong interaction. These particles and their correlations carry information on the properties of the medium in which they were created. The goal of all the existing and planned heavy ion experiments is to understand these properties and untangle the phases of matter from the particle information that is measured in the detectors. In many cases, new and interesting physics is hidden in rare events and/or rare particles as well as the correlations between these particles. Such is the case in the detection of new massive states such as nuclei, hypernuclei and metastable exotic objects as well as the properties of charmed hadrons and higher order cumulants of particle multiplicity distributions. 

To find and learn more about such rare probes, new experiments, like CBM  or PANDA at the upcoming FAIR facility, the RHIC beam energy scan , the NICA facility and the ALICE experiment at CERN, are designed to produce a huge amount of events every second. Since the amount of data generated in such events is very large, one has to find efficient methods to be able to classify and select new events very rapidly online in order to save them for later in-depth analysis. It is therefore desirable to have a model at hand which is able to quickly and reliably determine whether an event contains any potentially interesting information or is spoiled as compared to a background in some analysis. 

Another challenge with huge amounts of experimental data is related to possible interesting physics from a statistical analysis. A prominent example hereof is the analysis of the net-proton number multiplicity distribution as a function of beam energy by STAR at RHIC. In Au+Au collisions at $\sqrt{s}=7.7$~GeV, a significant deviation from a simple binomial distribution has been reported and interpreted as a signal for a critical endpoint in the QCD phase diagram \cite{Luo:2015ewa}, which is currently under intense investigation \cite{Herold:2018ptm,Herold:2017day,Herold:2016uvv,Nahrgang:2013jx}. As discussed in a previous work, this observation can, however, also be explained by a two peak anomaly in the proton number distribution \cite{Bzdak:2018uhv}. At the moment, the cause for such a two-bump distribution is unknown, possibilities include an experimental artifact or the effect of the QCD phase transition. 
To find the actual source, a careful analysis of the events responsible for that distribution may be useful which requires the identification of the corresponding events. If, for example, an imperfect centrality determination, a completely different event type, or even a detector malfunction would be responsible,  characteristics of those events should be different from those of the bulk. Such events are called outliers.

The detection of outliers has been an important branch in the machine-learning community (see e.g. \cite{hawkins}). In the present paper we will show how modern machine-learning (ML) methods can be applied to the detection of outliers in the context of high energy nuclear collisions. In particular, we will focus on Au+Au collisions at $\sqrt{s}=7.7$~GeV due to the previously mentioned observation of interesting fluctuations and correlations. Our work here can be understood as a suggestion for an extended experimental analysis of that particular beam energy.

In general, the presented methods are applicable to outlier detection in various nuclear collision experiments and are not at all restricted to the specific example discussed in the following. ML tools have nowadays become essential to face experimental challenges in high-energy physics \cite{Bourilkov:2019yoi,Radovic:2018dip,Guest:2018yhq,Larkoski:2017jix,deOliveira:2015xxd,Baldi:2016fql,Komiske:2016rsd,Almeida:2015jua,Kasieczka:2017nvn,Kasieczka:2019dbj,Qu:2019gqs,Moreno:2019bmu,Kasieczka:2020nyd,Sirunyan:2020lcu}, with applications ranging from track finding at PANDA \cite{Esmail:2019ypk} to b-jet tagging and the measurement of low-mass dielectrons at ALICE \cite{Haake:2017dpr}. Besides that machine learning has recently been used also in the study of heavy ion collisions \cite{Pang:2016vdc,Zhou:2018ill,Pang:2019aqb,Steinheimer:2019iso,Du:2019civ,Chien:2018rgm}.

The paper is structured as follows. First we will introduce the theoretical and model setup (section \ref{introduction}) and explain how the data for the machine-learning training process is generated (section \ref{sec:dataGeneration}). Then we will describe the ML models that are used for the task (section \ref{modelsForOutlierDetection}). Finally, we compare the performance of the different methods for the specific task and discuss results and implications for experiments (section \ref{results}).

\section{Generating data} \label{sec:dataGeneration}

In general, it is unknown how an outlier looks like, i.e. what its specific characteristics are and what distinguishes it from the bulk of events. We therefore  have to generate "unusual" events by hand and test our methods with these artificially created outliers. 

\begin{figure}[t]
    \centering
    \includegraphics[width=0.5\textwidth]{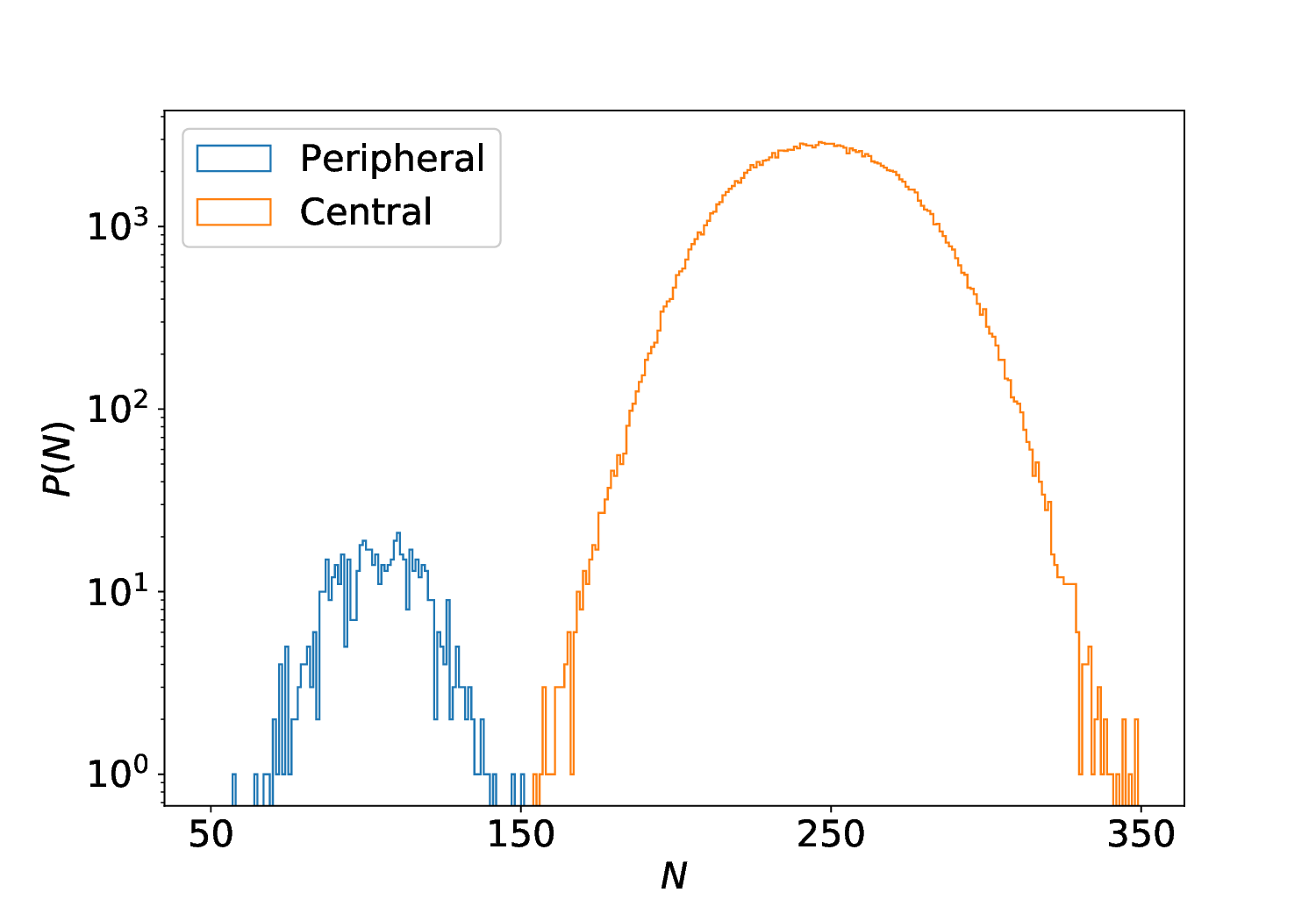}
    \caption{Distributions of the multiplicity of charged particles (excluding protons) $N$, using an acceptance cut of $|y|< 0.5$ and $P_T \leqslant \sqrt{2}$~GeV, in peripheral (blue) and central (orange) events. The two event classes show no overlap and are therefore are clearly distinguishable by their charged particle number.}
    \label{fig:pmd}
\end{figure}

For the study presented here, we generated 184000 central (with impact parameter $b=3$~fm) events and 600 peripheral ($b=7$~fm) events from Au+Au collisions at $\sqrt{s_{NN}} = 7.7$~GeV using the Ultrarelativistic Quantum Molecular Dynamics (UrQMD) transport model.
The UrQMD model is a simulation package based on an effective Monte Carlo solution of the Boltzmann transport equations \cite{Bass:1998ca, Bleicher:1999xi}. That means that hadrons are propagated on straight lines until they scatter according to experimentally known cross sections. Its application is widely used in high energy physics studies as it gives realistic results for the yields and momentum spectra of produced particles over a wide range of beam energies.

The reason for our specific choice of number of events for each centrality class is that such a combination of central (background) and peripheral (outlier) events would lead to an anomalous proton number distribution as it was observed by the STAR experiment. Our results can, however, be generalized to any number of events. 

Our goal is now to find a model which can successfully distinguish the outlier events (peripheral) from the large number of central events (background). 
Assuming a perfect experiment, such a choice of events would be easily distinguishable by the conventional method of counting the total number of charged particles per event \cite{Adamczyk:2013dal,Luo:2015ewa} as shown in figure~\ref{fig:pmd} where we show the charged particle distribution for peripheral and central events. It is obvious that the two have no overlap and therefore would allow for a clear separation. To manually make the events indistinguishable from such conventional analysis, we must remove the bias of the total number of charged particles. This can be done easily by e.g. normalizing the charged particle distributions, as we will explain later.

\begin{figure}[t]
    \centering
    \includegraphics[width=0.5\textwidth]{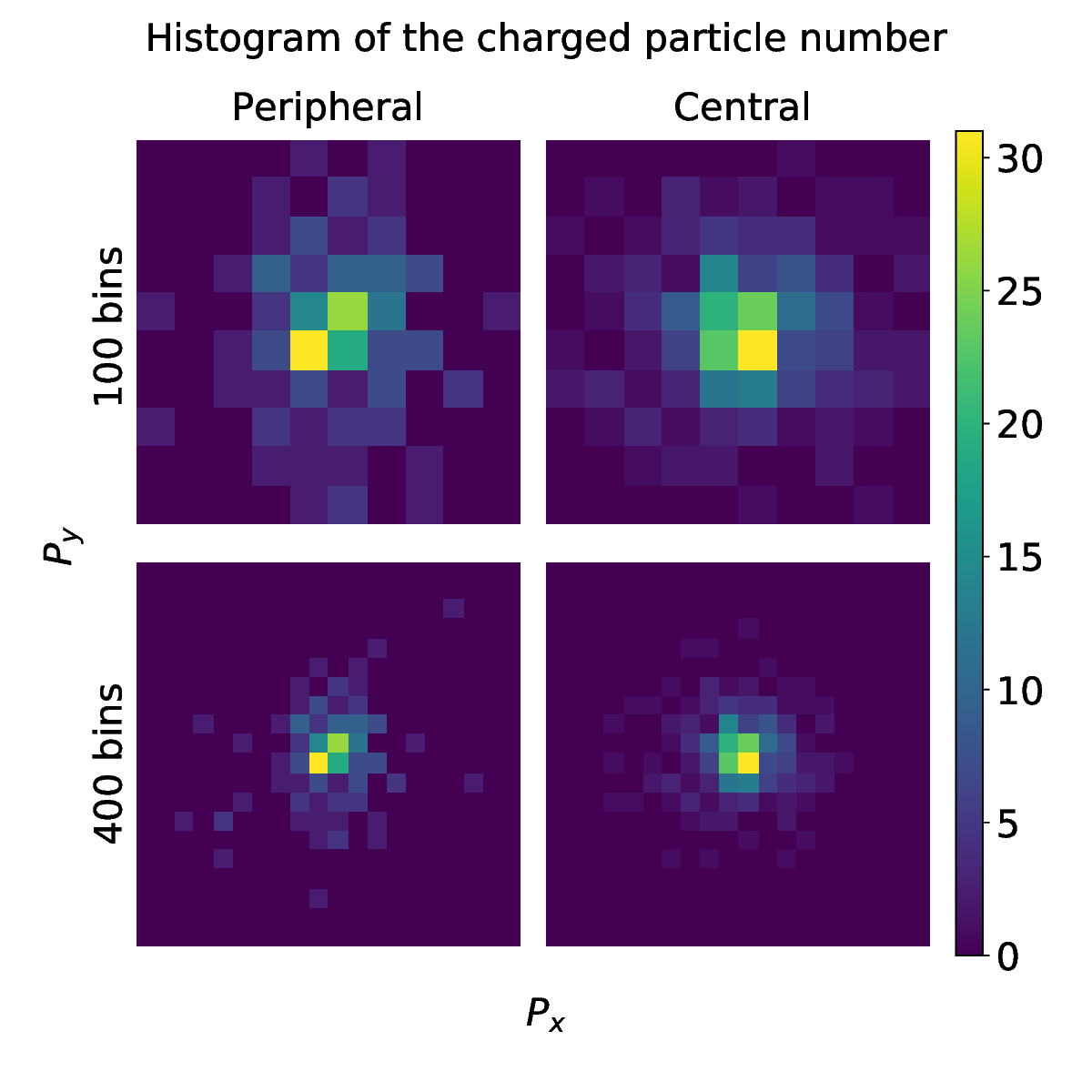}
    \caption{Several examples of the \textit{momentum features}, i.e. charged particle multiplicity distributions in the transverse momentum plane. Shown are examples for a single peripheral event (left column) and a central event (right column). The momentum ranges are $P_T \leqslant \sqrt{2}$~GeV,  divided into $10\times 10$ bins in the upper row, and $P_T \leqslant \sqrt{2}$~GeV, divided into $20\times 20$ bins in the lower row.}
    \label{fig:msp-w1w2}
\end{figure}

To use any outlier detection algorithm we first have to create characteristic features (or feature-vectors) for all individual events to serve as input for our outlier detection model. These features should be generated for each event using the same method and contain the characteristic information of that event. In the following we will use two dimensional histograms of the charged particle distributions in the X- and Y-momentum space as input features. Thus, in a first step, we create so-called \textit{momentum features} of charged particles by selecting all charged particles within the mid rapidity ($|y| \leqslant 0.5$) acceptance window. These particles are then binned in 2D-histograms of transverse $X$ and $Y$ momentum. The number of bins in each momentum-dimension is chosen to be either 10 or 20.

These \textit{momentum features} then characterize each individual event and form the basis of the training and validation datasets. The 2-dimensional histograms have the dimensionality of: 
\begin{itemize}
    \item $10 \times 10$ momentum bins along the $P_{X}$ and $P_{Y}$ direction, i.e. a total of 100 bins, for: \\
    $-1\  \mathrm{ GeV} \leqslant P_{X/Y} \leqslant 1 \  \mathrm{ GeV}$
    \item $20 \times 20$ momentum bins along the $P_{X}$ and $P_{Y}$ direction, i.e. a total of 400 bins, for: \\
    $-2\  \mathrm{ GeV} \leqslant P_{X/Y} \leqslant 2 \  \mathrm{ GeV}$
\end{itemize}

Examples of \textit{momentum features} for the two event classes are shown in figure~\ref{fig:msp-w1w2}. Here, one event of each centrality class is shown. The color code indicates the number of charged particles per $P_{X}$-$P_{Y}$ bin. As mentioned above, we intend to develop a method of event characterization which also works in cases where the total number of charged particles is an insufficient criterion or unusable. In a real experiment, this may occur if the standard separation method fails or is unreliable. We therefore have to make sure that such a simple way of separation is impossible for our ML outlier detection algorithm and do so by normalizing the 2D spectra (\textit{momentum features}) for each event by the number of charged particles in that event.

\section{Models for outlier detection} \label{modelsForOutlierDetection}

Outlier or anomaly detection is the task of detecting instances that deviate from a characteristic scenario or are an unlikely result of a random deviation from the expectation value. Outlier detection has a wide range of applications, e.g. e-mail spam detection, fraud detection for credit cards, etc. However, also in scientific fields and especially experimental data analysis of high energy physics, outlier detection has been used successfully \cite{Heimel:2018mkt,Farina:2018fyg}.
In this work, we present two kinds of algorithms for outlier detection, Principle Component Analysis (PCA) and Autoencoder Networks (AEN).
These algorithms can perform similar tasks and share some common features but also significant differences. While the PCA is a linear approach, the AEN can also take into account non-linear correlations in the training data.  In the following we will introduce the PCA and AEN in more detail and discuss their advantages and disadvantages as well as how they can be used. 

\subsection{Principle Component Analysis}

The Principle Component Analysis (PCA) is a statistical procedure that generates a low-dimensional representation of a dataset by an orthogonal linear transformation of the original data \cite{minka}. The input of the PCA must be a vector with dimension $n$. For our scenario this vector can be generated by flattening the input momentum feature. As a result we either obtain a 100 (from the $10\times 10$ histogram) or 400 (from the $20 \times 20$ histogram) dimensional vector. Note that the number of dimensions $n$ of the input is given by the number of independent bins in the histogram. Thus, for a $10 \times 10$ histogram, we have $n=100$.
The PCA then transforms the original $n$ coordinates of one input dataset into a new set of $m$ coordinates ($m<n$) called principal components (PC). These components are chosen to represent the data by maximizing the variance in the new set of $m$ dimensions.
The selection of principal axes can be understood as an iterative process. The PCA selects the first principal axis such that the variance of all datapoints in the direction of this new axis is maximized. The second axis is then selected to be orthogonal to the first one, again with the largest variance and so on. One can easily understand that the variance is a measure of the information content that is contained in a specific axis (dimension) selected as principal axis. The first PC therefore contains the largest amount of information, that can be used to characterize the input vector. As more principal components are added the cumulatively explained variance, i.e.  information stored in the PCA, increases (see also eq.~\eqref{eq:cev}). 

An important advantage of the PCA is that it can reduce the number of relevant features needed to largely reproduce the input. This also greatly improves the computational performance of the model. On the other hand the PCA is based on some assumptions which may not be optimal for every dataset, e.g. principal axis are orthogonal and that a large variance in a dimension implies more structure (even if that variance is due to noise).

In this work, we implement the PCA with a varying number of PCs to investigate the performance of the outlier detection as a function of the number of PCs.

To perform the PCA we use a freely available PYTHON machine learning library, Scikit-learn, and follow the procedure outlined in Ref.~\cite{minka}.

Having generated the original event-by-event momentum features and then used the PCA to reduce their dimensionality to $m$ dimensions, there are two methods available to find outliers based on the output of the PCA.

The first one is to make a so called \textit{radius comparison} ($RC$). After applying the PCA, each event is represented by an $m$-dimensional vector. The radius is defined as the length of each vector, in the new frame of principal components axes. It is calculated as:

\begin{equation}\label{eq:1}
    r(i) = \sqrt{X^2_{PC_1}(i) + X^2_{PC_2}(i) + .. + X^2_{PC_m}(i)}~,
\end{equation}

where $X_{PC}$ denote the $m$ coordinates of the new principal components and $i$ refers to the event number.

Using the PCA radius can help visualizing how outlier and background events are distributed in configuration space, the space where they are represented by principle components in a lower dimension, as we will see later.

The second method used to identify outliers is to calculate the \textit{reconstruction error} ($RE$) of a specific event. The reconstruction error quantifies the residuals induced when an input feature vector is projected on the new reduced set of $m$ dimensions and then is reconstructed from the reduced dimensions back to the original dimensionality. One could also say it quantifies the information loss that occurs for a single instance by performing the PCA.

We can calculate the reconstruction error of event $i$ simply as:

\begin{equation}\label{eq:2}
    RE(i) = \frac{1}{N}\left( \sum_{j=1}^{N}\left[ X_{rec_j}(i)-X_{j}(i)\right]^2\right)^{\frac{1}{2}}
\end{equation}
\noindent Hereby:
\begin{itemize}
\item $N$ is the number of dimensions of the input data (e.g., 100 or 400 for our momentum feature example).
\item {$X_{j}(i)$} is the $j^{th}$ component of the $i^{th}$ input event. Usually one sums over all $N$ components/dimensions of the input event.
\item {$X_{rec_j}$(i)} is the $j^{th}$ component of the of the $i^{th}$ reconstructed event. 
\end{itemize}

When we reduce the dimensionality of data while keeping a significant portion of the information in the principal components (defined as the components with the largest variance), the differences in the reconstruction loss are expected to differ between two types of events with different statistics (outliers and background). This is because the PCs are defined (`learned') by maximizing the variance, based on the background events. Therefore the same PC may not be ideally suitable to also maximize the variance for the outlier events. Thus, decomposing the outlier events using the principal components of the background may lead to a larger information loss for the outlier events. Therefore, the properties of the reconstruction error can be used as an indicator to detect an anomaly. 

\subsection{Autoencoder}\label{saen}

Autoencoders (AEN) are artificial neural networks that learn to reconstruct the input while changing the dimensionality of input data in a latent space. They follow a similar encoding-decoding strategy as PCAs but are not limited to linear projections of the input data, thus they can deal with more complex input. An AEN has been shown to also successfully denoise input data, which makes them very useful for the study of nuclear collision data. In an AEN, there are usually 3 components and 1 evaluator working together:

\begin{enumerate}
    \item The encoder is the network part that learns how to reduce the dimension of the input data and compresses it into an encoded representation.
    \item The Bottleneck (Hidden features, encoded representation,also called latent space) is the part that contains the dimensionally reduced representation of the input data. This part tries to preserve as much information as possible from the original input data.
    \item The Decoder is the part that decodes the bottleneck/hidden representation back to the output data, preserving as much significant information from the input data as possible. The output of the Decoder has the same dimensionality as the input data; in an ideal case, it is an almost perfect copy of the input.
    \item The Reconstruction Loss (reconstruction error) is an indicator that measures how well the output resembles the input and thus has been reconstructed from the hidden representation. The reconstruction loss can be defined in the same was as for the PCA shown in eq.~\eqref{eq:2}.
\end{enumerate}

One can apply many types of architectures such as fully-connected neural networks or convolutional neural networks for the encoder and decoder parts of the network depending on which structure can encode essential features of the input data most efficiently. These different architectures are explained in more detail in Appendix \ref{anns}. 

\begin{figure}[t]
    \centering
        \includegraphics[width=0.5\textwidth]{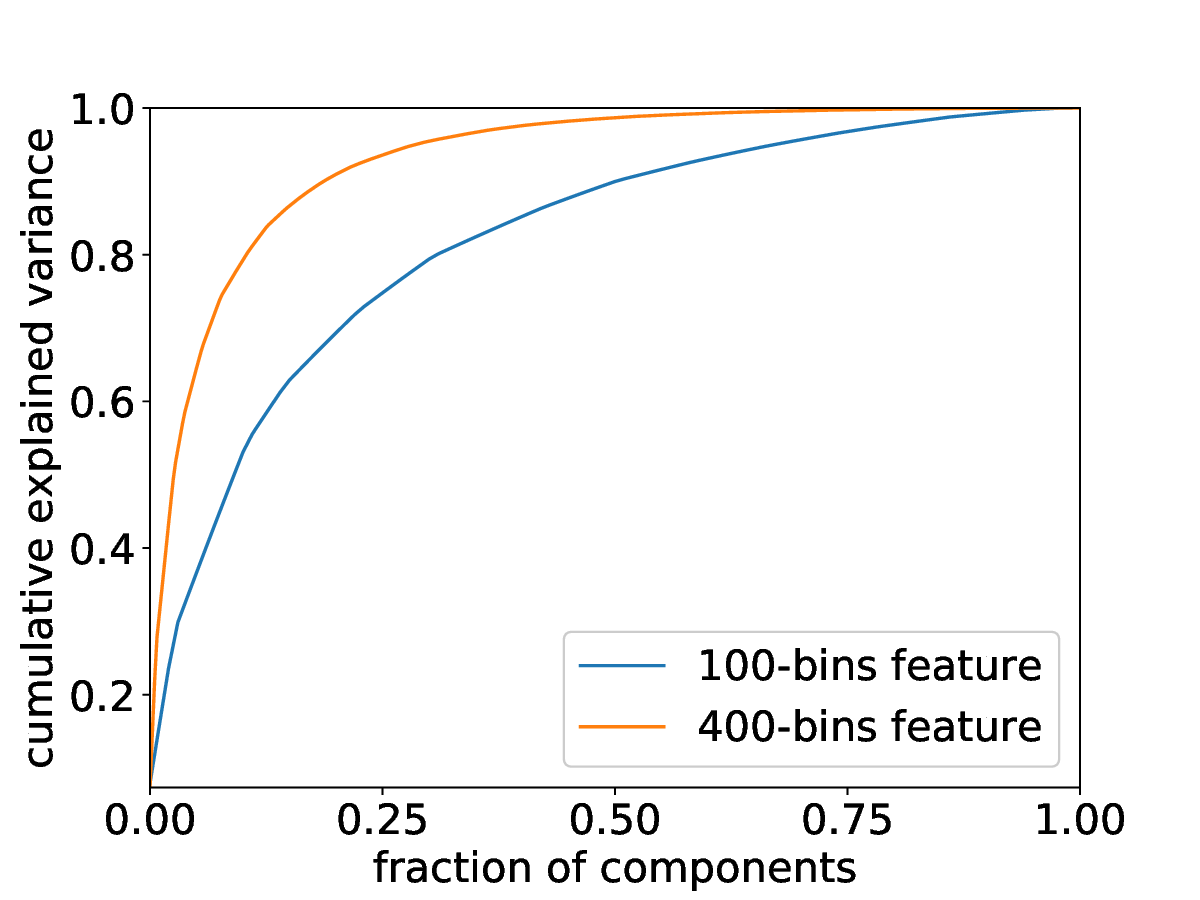}
\caption{The cumulative explained variance of the PCA as a function of the fraction of components. The fraction of components is equal to the number of PC used divided by the total number of input bins (i.e. 100 or 400 for the 100-bin and 400-bin features respectively). One can observe that with an increasing fraction of employed PCs, the variance of the data is explained better and thus the input is better reconstructed.}
\label{fig:cevw1w2}
\end{figure}

\section{Results} \label{results}

In this section, we present the results of implementing machine-learning algorithms on our generated data. The algorithms that we have described in the previous section, PCA radius comparison, PCA reconstruction error, and AEN reconstruction error, are applied and compared. We test the learning performance of each algorithm by varying parameters and applying it to different feature dimensions. 

\subsection{Radius comparison of PCA}

First, the PCA is used, with a varying number of PCs. To find out how much information is preserved with a given number of PCs we show the cumulative explained variance $\sigma_{cev}^2$ which is given by the cumulated variance of the $l$ used PCs divided by the cumulated variance of the maximum number $l_{max}$ of PCs which in our case is equal to 100 or 400, respectively. It reads 
\begin{equation}
\label{eq:cev}
    \sigma_{cev}^2 = \frac{\sigma_{l}^2}{\sigma_{l_{max}}^2} = \frac{\sum_{i=1}^{l} \sigma_{PC_i}^2}{\sum_{i=1}^{l_{max}}\sigma_{PC_i}^2}~. 
\end{equation}
Hereby, $\sigma_{PC_i}^2=\sum_{j=1}^{N} [(x_{PC_i})_j - \bar{x}_{PC_i}]^2/N$ is the variance of the $i$th PC over a total of $N=184600$ events. The cumulative explained variance is shown as function of the fraction of PCA  reduced dimensions $l/l_{max}$ in figure~\ref{fig:cevw1w2}. We see that for small values of the fraction of components, $\sigma_{cev}^2$ quickly rises. Several principal components are able to capture a significant part of the inputs variance. As we will see in the following, it is a priori not possible to predict whether more or less PCs will give a better separation of background and outlier events.

\begin{figure}[t]
    \centering
    \includegraphics[width=0.45\textwidth]{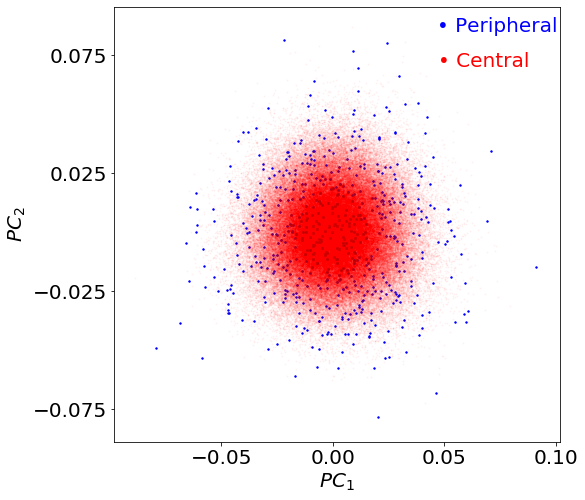}
    \caption{Representation of the input event after the PCA reduced the dimensionality of the input from 100  to 2 principle components. Each point represents one event. Peripheral (blue) and central (red) events are mixed, but the former ones are distributed more sparsely and spread over a larger area.}
    \label{fig:dplt2dw1}
\end{figure}

For illustrative purposes, we first investigate the distribution of the events in the case of using only 2 PCs. The distribution in coordinates $PC_1$ and $PC_2$ is shown in figure~\ref{fig:dplt2dw1}, showing background events as red dots and outlier events as blue dots. 
One can see that the distribution of both event classes is spherically symmetric, with peripheral events distributed more sparsely and over a larger area. There is, however, a significant overlap and it is clear that peripheral or signal events could only be uniquely identified if their radius, cf. eq.~\eqref{eq:1}, is sufficiently large. 

\begin{figure}[t]    
    \begin{subfigure}{}
        \includegraphics[width=0.5\textwidth]{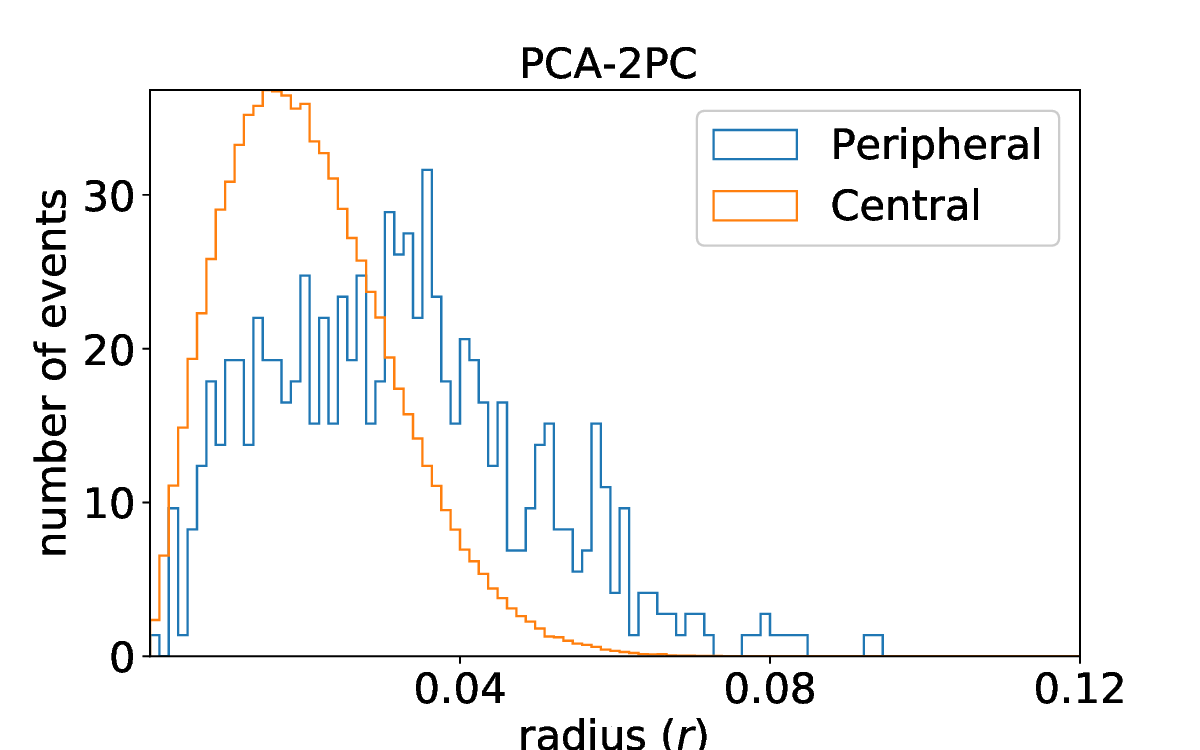}
         \includegraphics[width=0.5\textwidth]{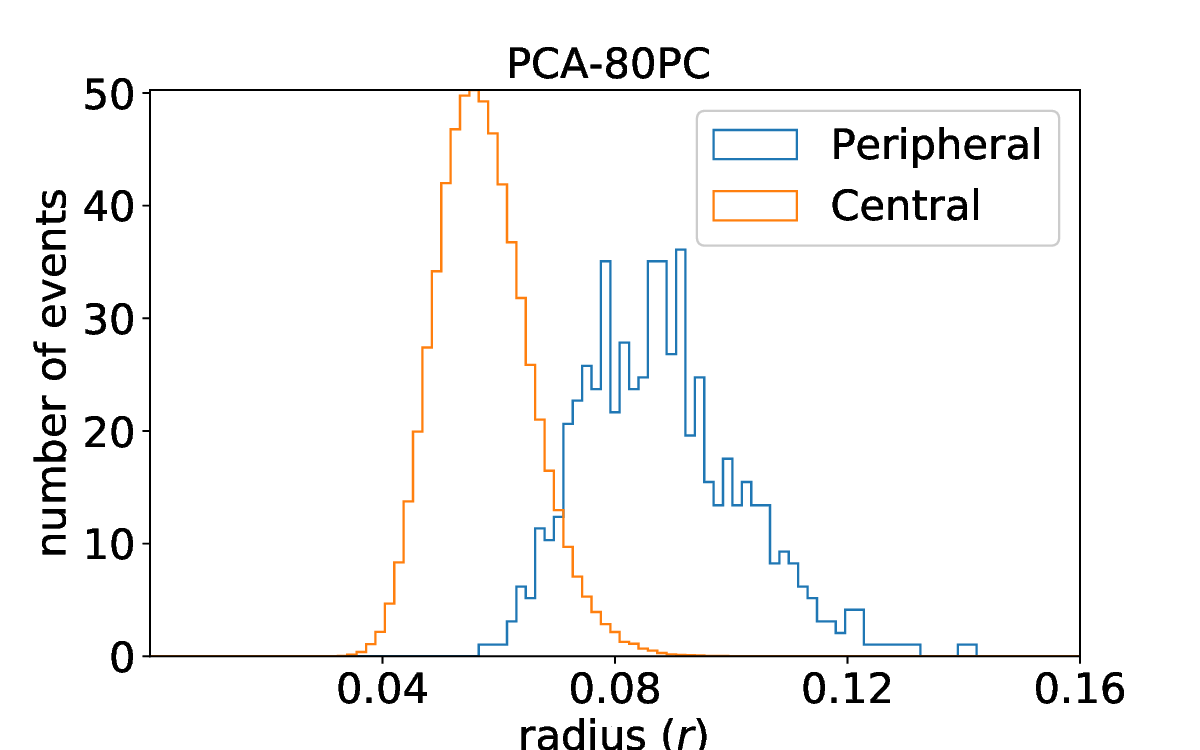}
    \end{subfigure}
    \caption{Distributions of the PCA radius for an input dimension of 100 bins, reduced to 2 PCs (top) and 80 PCs (bottom). We see that the higher number of PCs clearly yields a better separation of the two event classes.}
    \label{fig:pcarcw1}
\end{figure}

To demonstrate the actual overlap of the two event classes, histograms of the radii of the PCs are compared in figure~\ref{fig:pcarcw1}, for 2 and 80 PCs. The orange curves represent the distribution of the background while the blue curves represent the outliers.

\begin{figure}[t]
    \centering
    \includegraphics[width=0.48\textwidth]{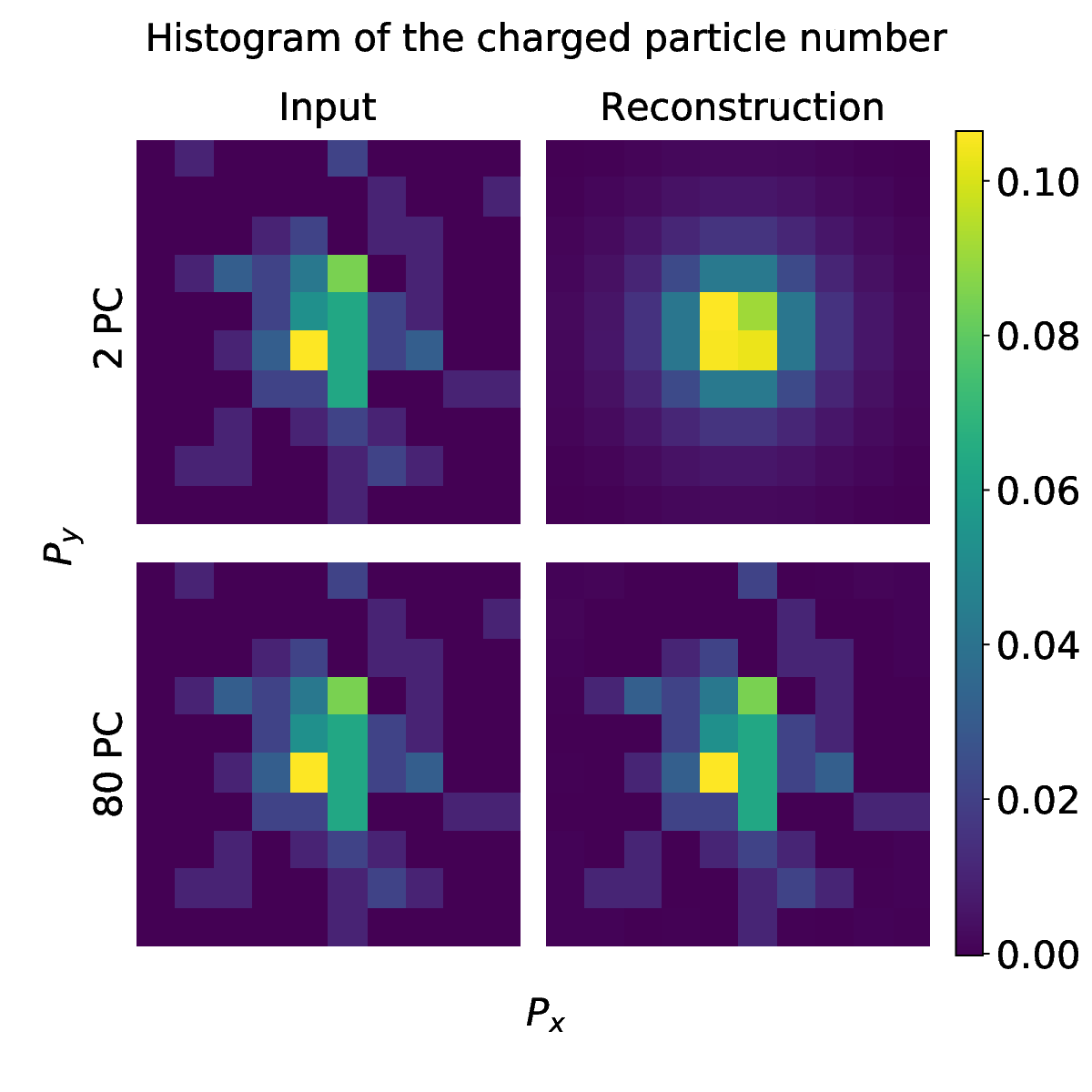}
    \caption{Comparison of the charged particle number input distributions with their reconstructions for a single event. The upper row corresponds to 2 PCs while the lower row corresponds to 80 PCs. For 2 PC, the reconstruction feature looks similar to the event-averaged input feature while for 80 PC, the reconstructed input is very similar to the input feature.}
    \label{fig:pcaeivo2d80dw1}
\end{figure}

While for $2$ PCs both distributions, signal and background, show a large overlap, the peaks are well separated in the case of $80$ PCs and the distributions share a much smaller overlap. It is obvious that, as more principal components are used, a better separation between the two classes can be achieved. To better understand this behaviour, we show a comparison of an arbitrary input event with its reconstruction by PCA using $2$ and $80$ PC in figure~\ref{fig:pcaeivo2d80dw1}. While the reconstruction with $2$ PCs essentially only recovers the average distribution, the reconstruction with $80$ PCs shows much more variability and closely resembles the input. This increased variability or variance is then captured in the radius, which is calculated as the sum of squares of all PCs, and therefore closely related to the variance. In other words, the PCs of the outlier events show a larger variance and more PCs make the identification of signal events easier.

\subsection{Reconstruction error of PCA}\label{ch:reop}

A different method to quantify the appearance of outlier events is to calculate the reconstruction error according to eq.~\eqref{eq:2}. Instead of focusing on the variability of the dataset as in the case of the radius comparison, this method focuses on the similarity of data points. 

\begin{figure}[t]
    \centering
     \begin{subfigure}{}
        \includegraphics[width=0.5\textwidth]{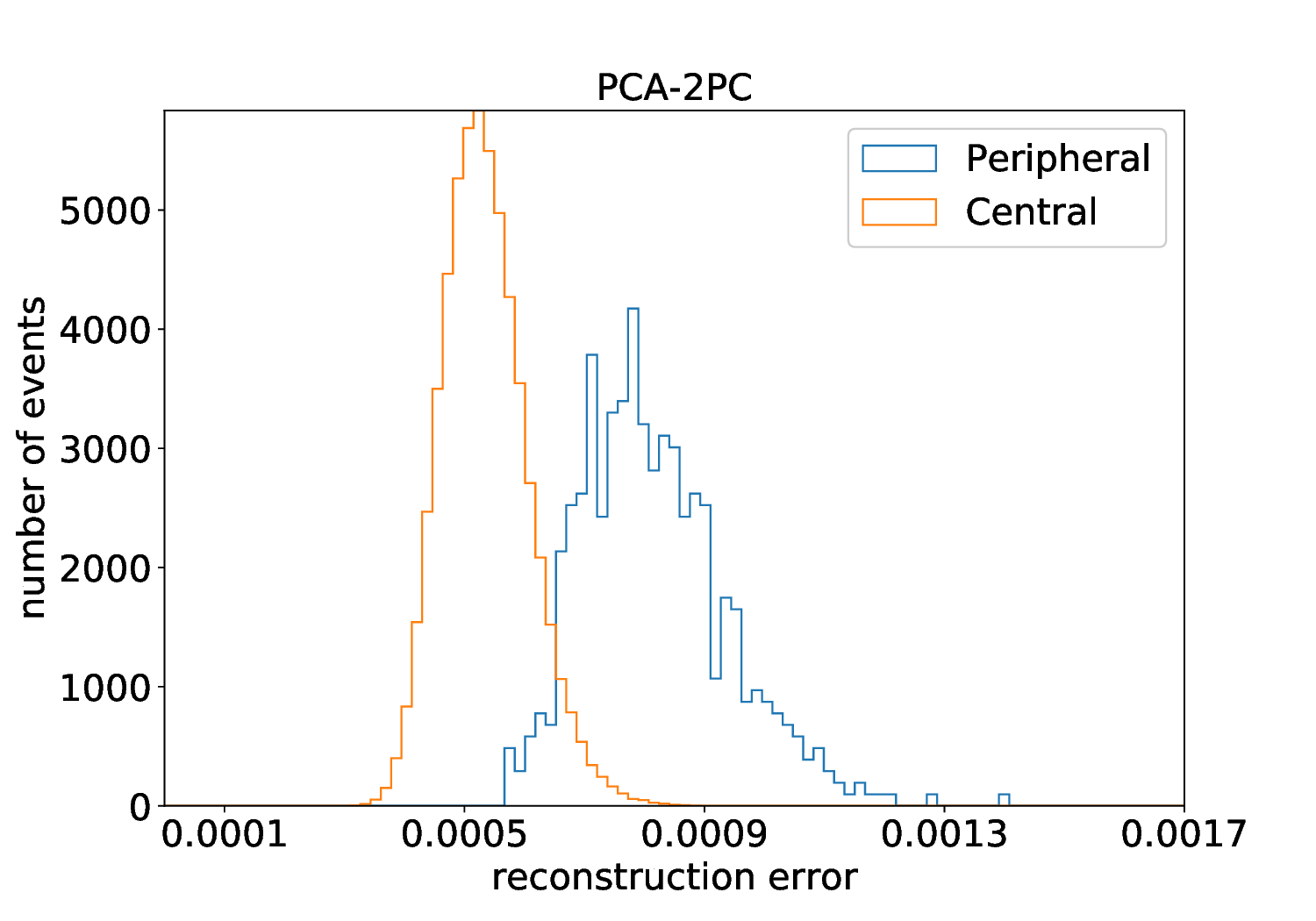}
        \includegraphics[width=0.5\textwidth]{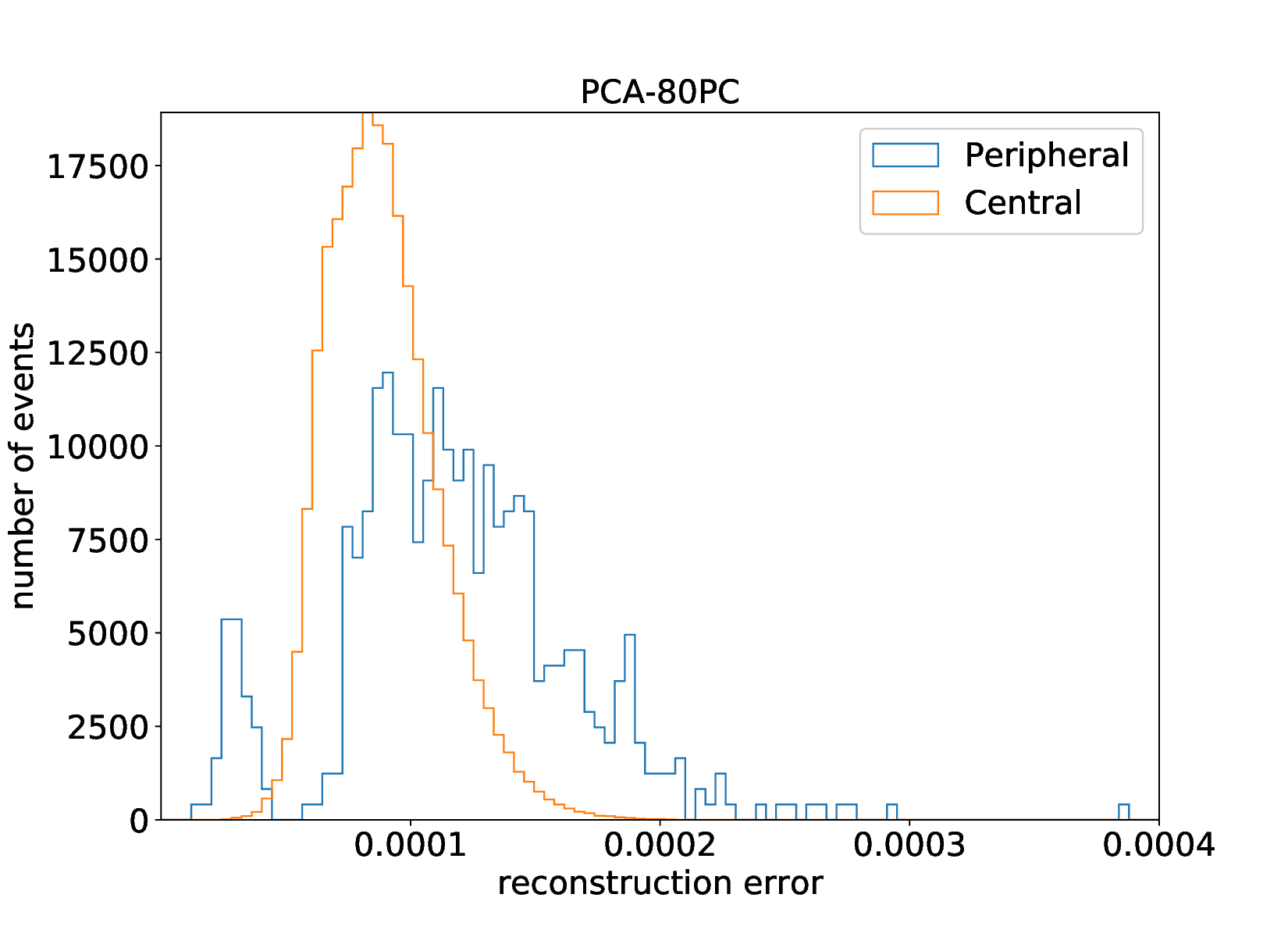}
    \end{subfigure}\\
    \caption{Comparison of the histograms of the reconstruction error for the PCA using 100 bin input features. The upper figure shows the results where only 2 PC are used, while the lower plot shows the reconstruction error for 80 PC. With a larger number of PCs, the reconstruction error decreases but also a larger overlap between the background and outliers is observed.}
    \label{fig:pcarew1}
\end{figure}
 
We calculate the reconstruction error for the cases $m=2$ and $m=80$ PCs. Figure~\ref{fig:pcarew1} shows the distributions of the reconstruction error for these two cases. We first note that the reconstruction error is overall much smaller for $80$ PCs than for $2$, as one would expect because the $80$ PCs carry much more information. On the other hand, the distributions in the case of $2$ PCs are better separated than for $80$. In the former case, background events are on average better reconstructed than outlier events. For $80$ PCs both background and outlier events are similarly well reconstructed as too much information is stored in the PCs. In that case both classes are not separable anymore using the $RE$.

The reconstruction error method is based on the similarity of events within a reduced representation and thus using less PCs works better to separate outlier and background events. In contrast to that, the radius comparison method differentiates the two classes by the variance of components, thus choosing a high number of PCs leads to a better discrimination. Therefore, the number of components needs to be chosen carefully and in accordance with the method to be used.

\subsection{ROC curve}

To quantify the quality of the different models' ability to find outliers, a \textit{Receiver Operating Characteristic} (ROC) curve is employed. 
The ROC curve is a method to estimate and compare performances of ML algorithms and it can tell how much of the background is considered to be part of the relevant outlier signals.

To create the ROC curve we start by dividing events into the learning algorithm predicted outliers and background, based on the distributions of the reconstruction error. To do so we define a threshold value of the reconstruction error above which all events are considered outliers. Since we do not have a unique way of determining the best threshold value, we simply calculate the fraction of correctly and incorrectly classified events for all possible threshold values. In practice, we gradually increase the threshold value of the reconstruction error and then count how many events above that value are actual outliers and how many belong to the background. As we increase the threshold, the fraction of signal over background will also increase steadily.

\begin{figure}[t]
    \centering
    \includegraphics[width=0.5\textwidth]{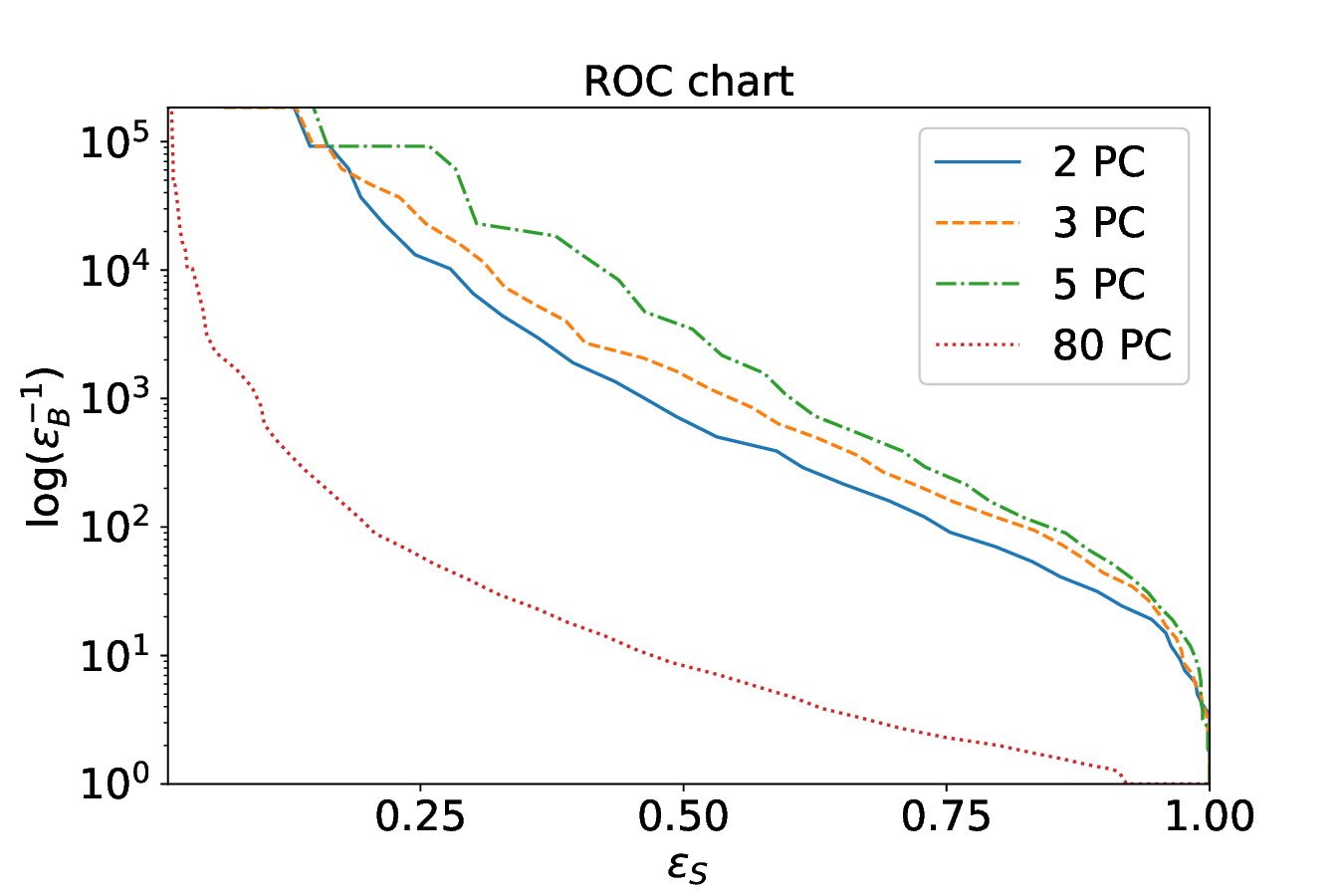}
    \caption{ROC chart to quantify the event separation using the PCA and the reconstruction error (RE). Compared are different numbers of PCs. We obtain the best result for 5 PC, corresponding to the highest curve.}
    \label{fig:roccpcare}
\end{figure}

In this work, we chose to plot the logarithm of the inverse of the fraction of incorrectly classified events $\log({\epsilon_b}^{-1})$ versus the fraction of correctly classified outlier events $\epsilon_{s}$, in the ROC curve. Here, the subscripts $b$ and $s$ stand for background and signal (outlier), respectively. Again, for a single threshold value we would obtain only one point in the ROC curve and the curve is generated by varying this threshold value. The choice of this threshold value then determines the amount of background which is overlapping with the outliers. Thus, the final choice of threshold will depend on the outcome of the ROC curve and how much background one is willing to accept while selecting out the outliers.

For our example with two centrality classes, peripheral (signal) and central (background), we define the fraction of events that are correctly classified as signal out of all actual signal events as 

\begin{equation}
    \label{eq:4}
    \epsilon_{s} = \frac{TP}{TP+FN}~.
\end{equation}
 
The fraction of events incorrectly classified as signal out of all actual background events is given by
 
\begin{equation}
    \label{eq:5}
    \epsilon_{b} = \frac{FP}{FP+TN}~.
\end{equation}
Hereby we use:
\begin{itemize}
    \item $TP$ (True Positives): number of correctly classified signal events
    \item $FN$ (False Negatives): number of wrongly classified background events
    \item $FP$ (False Positives): number of wrongly classified  signal events
    \item $TN$ (True Negatives): number of correctly classified background events
\end{itemize}

\begin{figure}[t]
    \centering
    \includegraphics[width=0.5\textwidth]{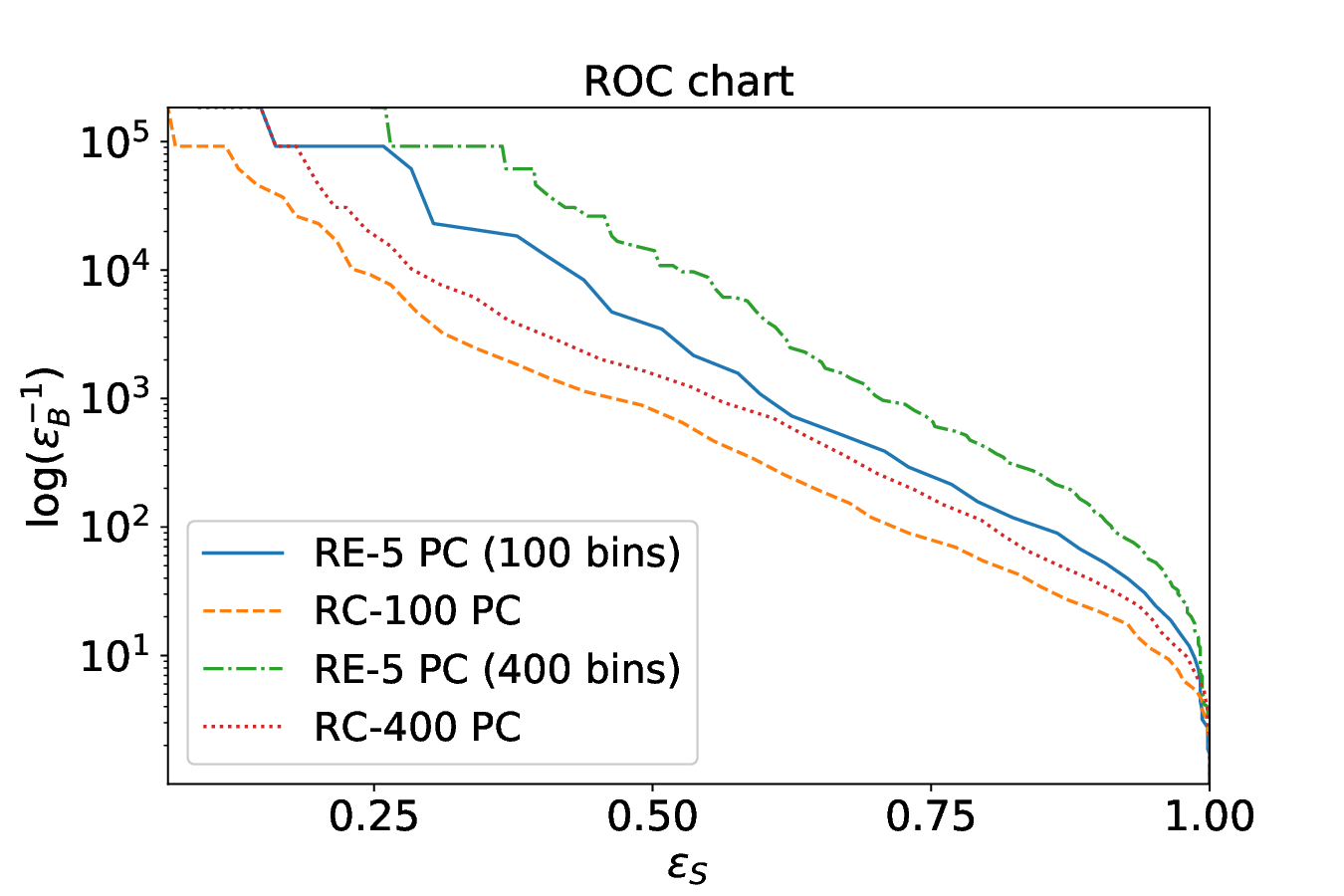}
    \caption{ROC chart comparing the two PCA methods, reconstruction error (RE) and radius comparison (RC). The solid lines correspond to the results with 100 input features and the dashed lines with 400 input features. All examples show the best performing outlier selector for each method respectively. This means that for the RC all input features are used and for the RE only 5 PC are used. We observe that the RE method yields a better performance for both input feature lengths.}
    \label{fig:pcarevsrc}
\end{figure}

According to eqs.~\eqref{eq:4} and \eqref{eq:5}, $\epsilon_s$ and $\epsilon_b$ are proportional to the number of correctly classified events and the number of wrongly classified events, respectively. Due to a large number of background compared to outlier events, a log scaling is applied to $1/\epsilon_b$. This makes the relation between $\epsilon_s$ and $\log({\epsilon_b}^{-1})$ more obvious to notice. 

The ROC curve can be interpreted in the following way: a large value of $\epsilon_s$ means that the threshold is chosen such that a large fraction of outlier events are selected. At the same time we want the fraction of background events selected with the same cut to remain small, thus $\log{\epsilon_b}^{-1}$ should be as large as possible. In other words, our ROC curve provides a clear performance measure in comparing different models, the best one determined as the one whose curve lies above all the others.

The ROC curves for the RE method using $m=1$, $3$, $5$, and $80$ PCs are shown in figure~\ref{fig:roccpcare}. Amongst these, $5$ PCs gives the best performance, clearly better than the large number of $80$ PCs but also significantly above the lower values of $3$ and $1$. Presumably, for $5$ PCs, enough significant information is preserved, while events are not reconstructed with too much details and consequently the outlier detection here works best. One should keep in mind that this specific choice of the number of PCs may be different for a different dataset and needs to be adjusted according to the specific experimental setup used.

\begin{figure}[t]
    \centering
     \begin{subfigure}{}
        \includegraphics[width=0.5\textwidth]{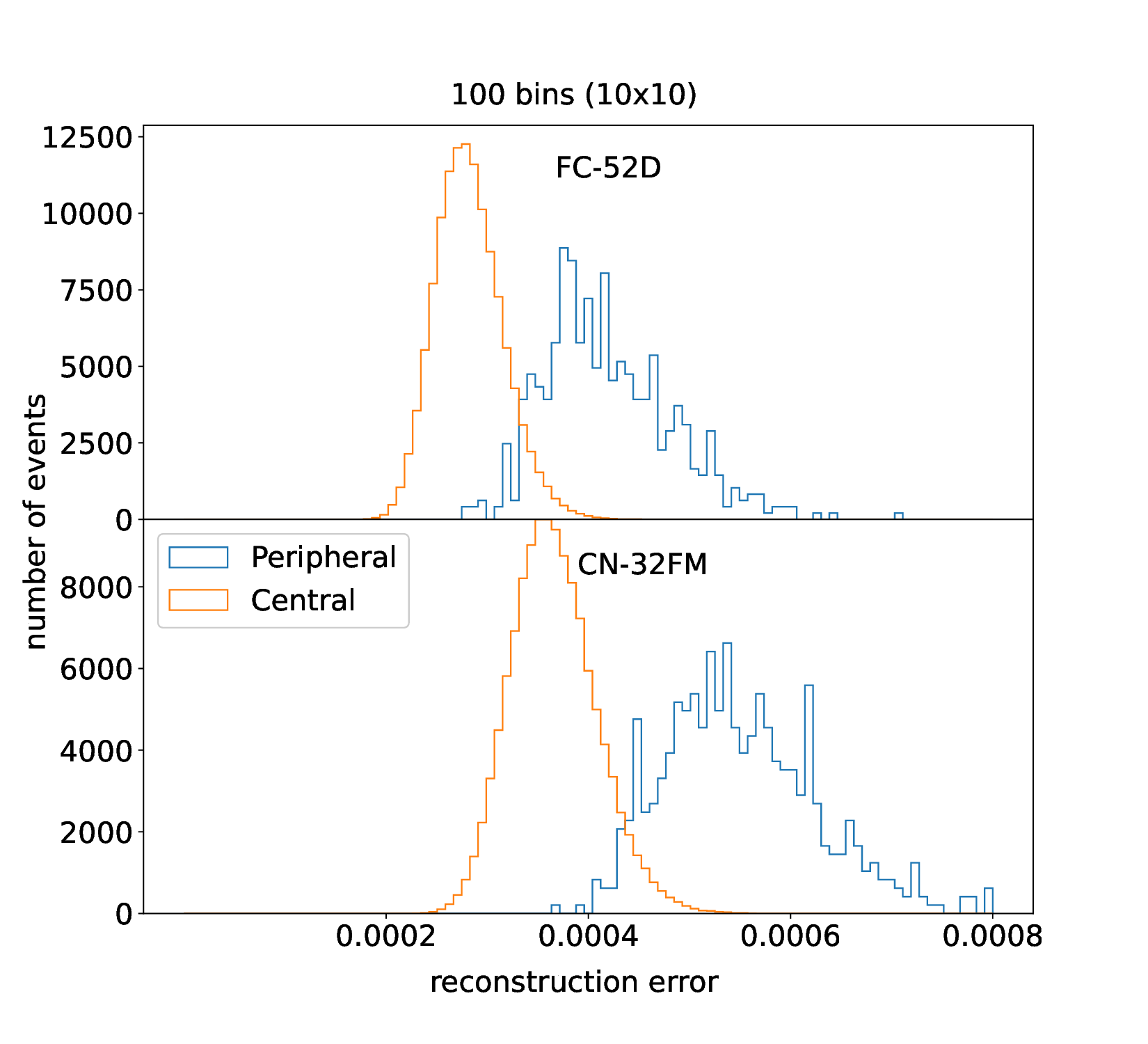}
    \end{subfigure}\\
    \caption{Comparison of the reconstruction error for the fully connected network with 52 neurons in the bottleneck (upper figure) and the convolutional neural network with 32 feature maps in the bottleneck (lower figure). The CNN shows the smallest reconstruction error but also the largest overlap between background (orange) and outlier (blue) events.}
    \label{fig:aenrew1}
\end{figure}

We also use ROC curves to compare the RC and RE methods using the PCA. 
The result is shown in figure~\ref{fig:pcarevsrc}. As one can see, the RE gives a better detection performance, i.e. for the same amount of detected outliers a smaller background is falsely included.

\subsection{Using an Autoencoder Network}

In this section we will explore the possibilities of using an Auto Encoder Network (AEN) for outlier detection in the same scenario as presented before.

Different kinds of neural network structures can be employed in the AEN. In the following, we use two different kinds of network structures, a fully-connected (FC) neural network and a convolutional neural network (CNN). The reason for choosing a convolutional neural network is that they are good at processing data in two dimensional array structures, such as the momentum feature maps we are using as input.
The same input features as used in the PCA, 100 and 400 bins, are used in the AEN study. The input features are encoded by the AEN to a lower dimensional representation at the networks bottleneck and subsequently decoded back to some output which has the same dimension as the input feature. The reconstruction error is then calculated according to eq.~\eqref{eq:2} for each event.

Unlike the PCA, neural networks can have an arbitrary size (number of neurons and layers) and even different structures like convolutional layer. This makes them a very powerful tool but also complicates the search for an appropriate structure for our task at hand. In many applications of neural networks the 'bigger is better' rule of thumb is a good starting point, so we will also first employ two different large networks with a large number of parameters. More details on the network structures used in the following are found in appendix \ref{anns}.

\begin{table}[b]
\centering
\begin{tabular}{|p{0.6cm}|p{2.0cm}||p{1.6cm}|p{2.0cm}||p{1.6cm}|}
 \hline
 & \multicolumn{2}{|c|}{10x10} & \multicolumn{2}{|c|}{20x20}\\
 \hline
 \multirow{3}{0.6 cm}{\rotatebox{90}{PCA}} 
                        & 2 PC & 3.5\% & 2 PC & 1.9\% \\
                        & 3 PC & 2.2\% & 3 PC & 1.1\% \\
                        & 5 PC & 1.6\% & 5 PC & 0.8\% \\
 \hline
 \multirow{3}{0.6 cm}{\rotatebox{90}{AEN}}
                        & FC-2D & 3.3\% & FC-2D & 1.8\% \\
                        & FC-3D & 2.5\% & FC-3D & 1.2\% \\
                        & FC-8D & 2.0\% & FC-8D & 0.9\% \\
 \hline
  \multirow{3}{0.6 cm}{\rotatebox{90}{AEN}}
                        & CN-1FM & 4.7\% & CN-1FM & 3.4\% \\
                        & CN-2FM & 3.6\% & CN-2FM & 8.1\% \\
                        & CN-4FM & 4.2\% & CN-4FM & 1.4\% \\
 \hline
\end{tabular}
\caption{This table summarizes the fraction of background (FP) events (in percent) are falsely identified as outliers if we cut on the reconstruction error to select 90$\%$ of true outliers (TP). Different numbers of encoded dimensions 2PC, 3PC, and 5PC for PCA and 2,3, and 8 for the fully connected AEN for both 10x10 and 20x20 momentum bins are compared to a varying number of feature maps for the CNN-AEN. All results using the input of 20x20 momentum bins separate the two classes better than the input of 10x10 momentum bins.
An optimal number of parameters is found for all models.}
\label{tab:pcaaentab1}
\end{table}

We first start with a fully connected network with 7 layers and a total of 35,976 trainable parameters. This network has 52 neurons in the bottleneck and the detailed network structure is shown in table \ref{tab:fc_5hl_st-w1} in the appendix. After training the network for 700 epochs on the input data, one obtains a reconstruction error of the order of $10^{-4}$ as can be seen in figure~\ref{fig:aenrew1}. However, as this model is very good at reconstructing the input, also the overlap between the two classes of outliers and background is significant, which leads to a large background rate of 3.3$\%$ if we want to extract $90\%$ of the outliers.
For the CNN the situation is similar. With 175,521 parameters the CNN AEN (shown in table \ref{tab:cn_st-w1} in the appendix) can achieve a reconstruction error of $10^{-4}$ as shown in figure~\ref{fig:aenrew1}. Again the very good reconstruction leads to a bad performance in outlier detection as the model is 'too good' at reconstructing any input.

To impair the network's reconstruction performance, we significantly reduce the number of layers and neurons in the bottleneck. In the following, we compare three different structures for the FC network and three for the CNN. For the FC network, $2$, $3$, and $8$ neurons are used in the bottleneck, and no other hidden layer is used. The structure of these networks is shown in table~\ref{tab:fc_st-w1} in the appendix. As one can see in table~\ref{tab:pcaaentab1}, the smallest bottlenecks also mix in a large amount of background with the outliers, they are too simple. As the number of neurons in the bottleneck is increased, the amount of background which is falsely identified as outlier decreases to a minimum. If we increase the number of neurons even further (larger than 8 for the FC network) the reconstruction error overall becomes small and a higher amount of background events are wrongly classified as outliers. In order to find an optimal network for outlier detection one therefore has to find the balance between reconstruction error and overlap with the background.

We also observe that in the case of a larger input dimensionality, the overlap between outliers and background is reduced which is an important finding.

Applying the CNN structure described in tables \ref{tab:cn_st-w1} and \ref{tab:cn_st-w2} in the appendix, and reducing the number of feature maps (FM), cf. Appendix~\ref{anns}, for each convolutional layer, again we observe an optimal number of FM. Here 2 or 4 FM are the optimal solution for the network structure employed, depending on the dimension of the input vector. Again, an optimal model means that it has the smallest fraction of background events which are falsely identified as outliers, in the case that $90 \% $ of all outliers are selected. 

Our results give some indication on how complex a network needs to be to be able to handle a certain input data and keep the balance between a good reconstruction but small overlap between background and outlier. Note that the result may change if different types of input data and dimensionalities are used.

\begin{figure}[t]
    \centering
    \includegraphics[width=0.48\textwidth]{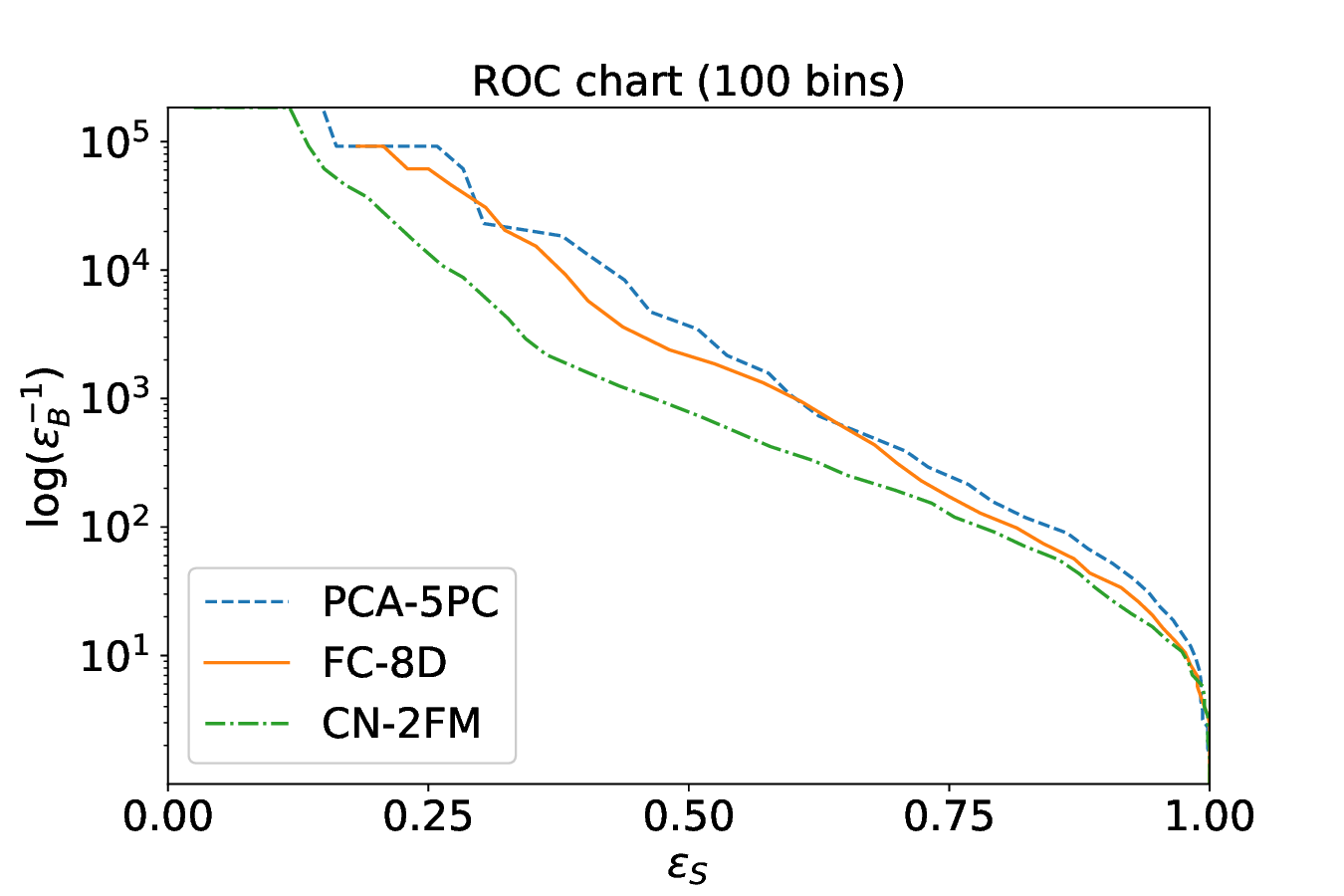}
    \caption{Comparison of the best models of each type: PCA, AEN-FC, AEN-CNN. The ROC chart shows that the performance of PCA and fully connected network are generally better than that of the CNN.}
    \label{fig:rocboerc}
\end{figure}

A final comparison of the performance of the PCA with two different autoencoder networks is presented in figure~\ref{fig:rocboerc} using ROC curves. Here, only the results for $10 \times 10$ momentum features are shown. We compare parametrizations which previously gave the smallest amount of falsely identified background for each model.
The PCA curve lies slightly above the other two, even though the difference with the FC network is marginal, indicating a better performance of the simplest model. The CNN generally yields the largest overlap between background and outliers. However, it also gives the smallest reconstruction error indicating that the CNN is too good at reconstructing the input even if the number of feature maps is reduced. This highlights the importance of understanding which network structure is suitable for which task. 

In practice, if a different input structure is used, different network structures may work better and need to be refined. Nevertheless, in the following we will use the three models from figure~\ref{fig:rocboerc} for identifying different types of outliers, employing the best structure found for each. This will give us some indication on the robustness of the structures.

\begin{figure}[t]
    \centering
     \begin{subfigure}{}
        \includegraphics[width=0.45\textwidth]{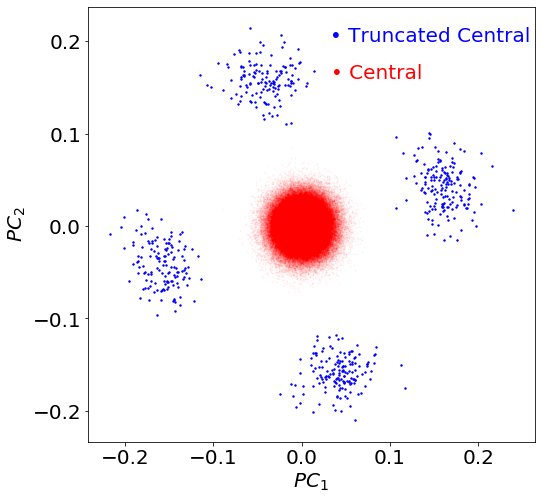}
    \end{subfigure}\\
    \caption{The 2-dimensional representations of events where one quadrant of the input feature was removed. The PCA was applied to reduce the dimensionality to 2 PCs. The figure shows that the outliers (blue) are now well separated from the background (red) as 4 groups. Each group corresponds to one removed quadrant in the input feature.}
    \label{fig:ch2-pcadpw}
\end{figure}

\subsection{Testing with different outlier types}

After training and selecting our models in separating central and peripheral event classes, we now test them on different types of outliers which again have to be constructed artificially. We create three different types of outlier classes as follows: We generate 600 central events and their corresponding momentum features as detailed earlier in section~\ref{sec:dataGeneration}.

\begin{itemize}
    \item Option I: We then truncate one quarter of these features, i.e.~we set the values in one randomly chosen quadrant of the $P_X$-$P_Y$ plane equal to zero. 
    \item Option II: We truncate $20$\% of the $P_X$ and $P_Y$ range, leaving only the inner $8\times 8$ or $16\times 16$ bins of the 100- or 400-bins features unaltered. I.e. we cut all high momentum features.
    \item Option III: With a probability of $10$\%, we randomly set the value of any given momentum bin to zero. This will create non-Gaussian noise in the truncated events.
\end{itemize}

All obtained spectra are again normalized to $1$. We do this with both 100-bin and 400-bin features. 

We begin with the most severe outliers of option I. Here we first check the 2-dimensional representation in PCs after applying the PCA, cf. figure~\ref{fig:ch2-pcadpw}. While the central events naturally occupy the same space as shown earlier in figure~\ref{fig:dplt2dw1}, the truncated events or outliers are clearly separated from these into four distinct areas which are symmetrically positioned around the central events. Each of four these areas contains outlier events where one specific quadrant has been set to zero. 

Such outliers are very easy to separate for the PCA as well as for the AEN. In fact for all models, we obtain a zero percent overlap if 90$\%$ of the outliers are selected. Such a severe malfunction of the detector can therefore be easily and readily detected by a simple outlier detection algorithm.

Table~\ref{tab:pcaaentab2} compares the performance of all models when tested on the different test cases. In the scenarios II and III, the obtained overlap is larger than for our training scenario where peripheral events were used as outliers. Obviously, the detection of outliers becomes more difficult when only a small percentage of the information is inaccessible. Again, using higher dimensional input features seems to improve the models' ability to distinguish outliers from background.

\begin{table}[b]
\centering
\begin{tabular}{|p{1.5cm}|p{1.0cm}|p{1.0cm}|p{1.0cm}|p{1.0cm}|p{1.0cm}|p{1.0cm}|}
 \hline
  \multirow{2}{0.8 cm}{Input dim.} & \multicolumn{2}{|c|}{PCA} & \multicolumn{2}{|c|}{AEN-FC} & \multicolumn{2}{|c|}{AEN-CN} \\
             & 100  & 400  & 100  & 400  & 10x10 & 20x20\\
            \hline
\multirow{2}{1.5 cm}{Option I} & 5PC & 5PC & 5D & 5D & 2FM & 4FM \\
                               & 0\% & 0\% & 0\% & 0\% & 0\% & 0\% \\
 \hline
 \multirow{2}{1.5 cm}{Option II} & 5PC & 5PC & 5D & 5D & 2FM & 4FM  \\
                               & 2.5\% & 1.4\% & 2.5\% & 1.4\% & 3.3\% & 2.8\% \\
 \hline
 \multirow{2}{1.5 cm}{Option III} & 5PC & 5PC & 5D & 5D & 2FM & 4FM  \\
                               & 4.3\% & 1.0\% & 5.0\% & 1.1\% & 12.2\% & 1.5\% \\
 \hline

\end{tabular}
\caption{Comparison of the test results for the three different models. Here the fraction of falsely as outlier identified background is shown for the three outlier options that are used in the testing. Both PCA and the AEN-FC perform equally well.}
\label{tab:pcaaentab2}

\end{table}

\section{Conclusion} \label{conclusion}

We have presented and explored methods of outlier detection using unsupervised learning which can prove useful for data analysis in high energy nuclear collision experiments. For this purpose, we have compared several unsupervised machine learning models such as the Principal Component Analysis (PCA) and Autoencoder networks (AEN). In a specific example we use the unsupervised learning to separate misidentified peripheral events from a background of central events. This example was motivated by the yet unexplained finding of large factorial cumulants at the STAR experiment. In this specific example, the transverse momentum spectra served as input features for the ML algorithms. 

It was found that the reconstruction error in PCA or AEN can be a useful tool to identify outlier events. Furthermore, using the reconstruction error, it was found that a model which is too complex, i.e. gives a very small reconstruction error, gives a larger overlap of background and outlier events. Thus, a model which is less complex (has fewer parameters), but complex enough to capture the most essential features of the event is preferred. This is consistent with the result that a higher dimensional input feature also provides a better separation capability, as it is harder to reconstruct exactly but general features can be captured also by less complex models. This therefore provides an advantage for the direct application of outlier detection in an online analysis tool for heavy ion experiments. Since the model can be less complex it will be able to handle more events in a shorter time with fewer computational resources.

In a practical application the methods presented here will have to be adjusted to the actual output of the experiment. However, we believe that our work can provide a solid guideline for the application of unsupervised outlier detection in nuclear collision experiments.

\acknowledgments
This work was supported by the Development and Promotion of Science and Technology Talents Project (DPST) -
Royal Thai Government Scholarship, Suranaree University of Technology (SUT). JS and KZ thank the Samson AG and the BMBF through the ErUM-Data project for funding. JS thanks the Walter Greiner Gesellschaft zur F\"{o}rderung der physikalischen Grundlagenforschung e.V. for its support. KZ gratefully acknowledges supports from the
NVIDIA Corporation with the GPU Grant. This work was supported by the DAAD through a PPP exchange grant. Computational resources were provided by the Frankfurt Center for Scientific Computing (Goethe-HLR).

\appendix
\section{Artificial Neural Networks (ANNs)} \label{anns}

Artificial neural networks are essentially mapping functions that map an $n$-dimensional input on an $m$-dimensional output. In the specific case of an autoencoder network, the input dimension is equal to the output dimension. Besides these so-called input and output layers a neural networks consists of a varying number of hidden layers, with neurons who themselves perform a non-linear transformation on their input. The output $y=a x +b$ of any given neuron (where $a$ and $b$ are parameters and $x$ is the input of the neuron) serves as argument of a so-called \textit{activation function} which can take on different forms (often sigmoid or Relu functions are used). Depending on the structure of the network, the neurons of one single layer can be connected to any number of neurons in the next layer. For example, in a fully connected neural network, the output of the $j^{\mathrm{th}}$ neuron in the $(i+1)^{\mathrm{th}}$ layer is the sum of all outputs of the $i$th layer $y_{i+1,j}=f(\sum_{k} a_{k,i+1} x_k +b)$ where $k$ is the index of a neuron in the $i^{\mathrm{th}}$ layer and $f()$ is an activation function. Such a network can easily have a large number of parameters $a_{j,i}$ and $b_{j,i}$ to be determined. The determination of these parameters is done during the training phase of the network where a loss function is minimized. In the autoencoder network, the loss function is simply defined as the reconstruction error of the input vs. output comparison. In our specific case this is done by calculating the mean squared error ($mse$) of the networks output with respect to the networks input, where the squared errors for all dimensions are summed. The parameter values of the network are changed using a gradient descent method in order to minimize the error.
The gradient descent in our calculations is done using the Adam optimizer ($adam$) which is provided by the \textit{Tensorflow} library. For a much more in-depth explanation on neural networks we refer the interested reader to \cite{Mehta:2018dln}

\subsection{Fully Connected Neural Networks}
The term fully connected comes from the fact that all nodes in one layer of a network are connected to the nodes of the next layer. Fully connected neural networks are good at pattern recognition. 
For the results presented, the following fully connected structure have been used: (note that The number in parenthesis indicate the number of neurons used for each layer which is equal to the number of outputs in the case of a fully-connected neural network)\\

\begin{table}[h!]
\centering
\begin{tabular}{|p{1.0cm}|p{1.4cm}|p{1.4cm}|p{1.5cm}|p{1.4cm}|}
 \hline
 \multicolumn{5}{|c|}{FC (100 bins), Opt='adam', Loss='mse'}\\
 \hline \hline
  layer & types & nodes & act fn & output\\
 \hline
     1 & Input & 100 &  & 100\\
 \hline
     2 & Dense & 88 & relu & 88\\
 \hline
     3 & Dense & 64 & relu & 64\\
 \hline
     4* & Dense & 52 & relu & 52\\
 \hline
     5 & Dense & 64 & relu & 64\\
 \hline
     6 & Dense & 88 & relu & 88\\
 \hline
     7 & Dense & 100 & sigmoid & 100\\
 \hline
\end{tabular}
\caption{Structure of the largest (with respect to the number of parameters) fully connected network used. Shown is the type of layer, the number of neurons in that layer (nodes), the activation function used for that layer and the output dimensionality of that layer. The total number of parameters of this model is 35,976.}
\label{tab:fc_5hl_st-w1}
\end{table}

\begin{table}[h!]
\centering
\begin{tabular}{|p{1.0cm}|p{1.4cm}|p{1.4cm}|p{1.5cm}|p{1.4cm}|}
 \hline
 \multicolumn{5}{|c|}{FC (100 bins), Opt='adam', Loss='mse'}\\
 \hline \hline
  layer & types & nodes & act fn & output\\
 \hline
     1 & Input & 100 &  & 100\\
 \hline
     2* & Dense & 2/3/5 & sigmoid & 2/3/5\\
 \hline
     3 & Dense & 100 & linear & 100\\
 \hline
\end{tabular}
\caption{Structure of the smaller (with respect to the number of parameters) fully connected networks used with the $10\times10$ dimensional input features. Shown is the type of layer, the number of neurons in that layer (nodes, which is either 2, 3 or 5), the activation function used for that layer and the output dimensionality of that layer. The total number of parameters of this model are 502/703/1,105 for 2/3/5 encoded dimensions in the bottleneck respectively.}
\label{tab:fc_st-w1}
\end{table}

\begin{table}[h!]
\centering
\begin{tabular}{|p{1.0cm}|p{1.4cm}|p{1.4cm}|p{1.5cm}|p{1.4cm}|}
 \hline
 \multicolumn{5}{|c|}{FC (400 bins), Opt='adam', Loss='mse'}\\
 \hline \hline
  layer & types & nodes & act fn & output\\
 \hline
     1 & Input & 400 &  & 400\\
 \hline
     2* & Dense & 2/3/5 & sigmoid & 2/3/5\\
 \hline
     3 & Dense & 400 & linear & 400\\
 \hline
\end{tabular}
\caption{Structure of the smaller (with respect to the number of parameters) fully connected networks used with the $20\times20$ dimensional input features. Shown is the type of layer, the number of neurons in that layer (nodes, which is either 2, 3 or 5), the activation function used for that layer and the output dimensionality of that layer. The total number of parameters of this model are 2,002/2,803/4,405 for 2/3/5 encoded dimensions in the bottleneck respectively.}
\label{tab:fc_st-w2}
\end{table}

\subsection{Convolutional Neural Networks}

 Convolutional neural networks are a special kind of network structure which takes into account the two dimensional structure of the input images. The CNN takes a two dimensional array as input as well as output. Instead of mapping every input pixel to an independent neuron the CNN uses convolutional kernels on the input image. These convolutional kernels have the dimension $m \times m$ or written $(m,m)$. In each convolutional layer these kernels can take the form of $n$ so called feature maps (FM) which constitute the trainable parameters of the model. The general training procedure is similar to that of the fully connected network in that a gradient descent algorithm is used to change the parameters such that the loss function (again the mean squared error) is minimized.
 In addition to the kernel size and the feature maps, the CNN also can use strides (step size with which the kernels scan the image) and padding (additional pixels which are added on the images boundary).
 In the following the CNN structures used in the result section are summarized.

\begin{table}[h!]
\centering
\begin{tabular}{|p{0.8cm}|p{2.0cm}|p{1.1cm}|p{1.3cm}|p{1.25cm}|p{1.4cm}|}
 \hline
 \multicolumn{6}{|c|}{CN (10x10), Opt='adam', Loss='mse', 700 epochs}\\
 \hline
  layer & FM, kernel & strides & padding & act fn & output\\
 \hline
     1 &  &  &  &  & (10,10,1)\\
 \hline
     2 & 4/8/16/128, (2,2) & (2,2) & same & relu & (5,5,4)\\
 \hline
     3 & 2/4/8/64, (3,3) & (2,2) & same & relu & (3,3,2)\\
 \hline
     4* & 1/2/4/32, (3,3) & (2,2) & same & sigmoid & (2,2,1)\\
 \hline
     5 & 2/4/8/64, (2,2)$^{T}$ &  &  & relu & (3,3,2)\\
 \hline
     6 & 4/8/16/128, (3,3)$^{T}$ &  &  & relu & (5,5,4)\\
 \hline
     7 & 1, (2,2)$^{T}$ & (2,2) & same & relu & (10,10,1)\\
 \hline
\end{tabular}
\caption{Structure of the convolutional AEN used with the $10 \times 10$ dimensional input features. Shown is the number of feature maps (FM) and the kernel size in that layer, the strides, the activation function used and the output dimensionality of that layer. The total number of parameters of this model are 216/771/2,901/175,521 for 1/2/4/32 FMs in the bottleneck respectively.}
\label{tab:cn_st-w1}
\end{table}

\begin{table}[h!]
\centering
\begin{tabular}{|p{0.8cm}|p{2.0cm}|p{1.1cm}|p{1.3cm}|p{1.25cm}|p{1.4cm}|}
 \hline
 \multicolumn{6}{|c|}{CN (20x20), Opt='adam', Loss='mse', 700 epochs}\\
 \hline
  layer & FM, kernel & strides & padding & act fn & output\\
 \hline
     1 &  &  &  &  & (20,20,1)\\
 \hline
     2 & 4/8/16, (2,2) & (2,2) & same & relu & (10,10,4)\\
 \hline
     3 & 2/4/8, (2,2) & (2,2) & same & relu & (5,5,2)\\
 \hline
     4 & 2/4/8, (3,3) & (2,2) & same & relu & (3,3,2)\\
 \hline
     5* & 1/2/4, (3,3) & (2,2) & same & sigmoid & (2,2,1)\\
 \hline
     6 & 2/4/8, (2,2)$^{T}$ &  &  & relu & (3,3,2)\\
 \hline
     7 & 2/4/8, (3,3)$^{T}$ &  &  & relu & (5,5,2)\\
 \hline
     8 & 4/8/16, (2,2)$^{T}$ & (2,2) & same & relu & (10,10,4)\\
 \hline
     9 & 1,(2,2)$^{T}$ & (2,2) & same & relu & (20,20,1)\\
 \hline
\end{tabular}
\caption{Structure of the convolutional AEN used with the $20 \times 20$ dimensional input features. Shown is the number of feature maps (FM) and the kernel size in that layer, the strides, the activation function used and the output dimensionality of that layer. The total number of parameters of this model are 212/747/2,789 for 1/2/4 FMs in the bottleneck respectively.}
\label{tab:cn_st-w2}
\end{table}

\section{Computational Libraries Used} \label{mlLibraries}
To construct and train the networks used in this work different numerical PYTHON libraries are used:

\begin{itemize}
    \item Scikit-learn: an open source machine learning library that used to perform the PCA \cite{minka}.
    \item TensorFlow: an open source machine learning library that support artificial neural network construction and training.
    \item Keras: a library that uses tensorflow as backend.
\end{itemize}

\newpage

\newpage

\bibliographystyle{JHEP}

\end{document}